# Optimal Design of Energy-Efficient Millimeter Wave Hybrid Transceivers for Wireless Backhaul

Andrea Pizzo and Luca Sanguinetti

Dipartimento di Ingegneria dell'Informazione, University of Pisa, Pisa, Italy

*Abstract*—This work analyzes a mmWave single-cell network, which comprises a macro base station (BS) and an overlaid tier of small-cell BSs using a wireless backhaul for data traffic. We look for the optimal number of antennas at both BS and small-cell BSs that maximize the energy efficiency (EE) of the system when a hybrid transceiver architecture is employed. Closed-form expressions for the EE-optimal values of the number of antennas are derived that provide valuable insights into the interplay between the optimization variables and hardware characteristics. Numerical and analytical results show that the maximal EE is achieved by a 'close-to' fully-digital system wherein the number of BS antennas is approximately equal to the number of served small cells.

## I. INTRODUCTION

Millimeter Wave (mmWave) communications suffer from high atmospheric absorption, rain and foliage attenuation, penetration and reflection losses, which essentially restrict their use to line-of-sight (LoS) indoor-to-indoor or outdoor-to-outdoor communications over relatively short distances [1]. Nevertheless, recent theoretical considerations and measurement campaigns have provided evidence that outdoor cells with up to 200 m cell radii are viable if transmitters and receivers are equipped with sufficiently large antenna arrays along with beamforming [2], [3]. However, large arrays beamforming poses several implementation challenges mostly because of hardware limitations that make hard to have a dedicated baseband and radio frequency (RF) chain for each antenna. Analog solutions arise in early works for mmWave systems for their ease of implementation and power saving [4], [5] at the price of single-stream transmissions that substantially limit the system spectral efficiency. To combine the benefits of analog and digital architectures, hybrid beamforming schemes have gained a lot of interest [6].

A hybrid beamformer is made up of a low-dimensional baseband precoder followed by a high-dimensional RF beamformer. The latter is fully implemented by low-cost and power efficient analog phase shifters. Interestingly, in [7] the authors provide necessary and sufficient conditions to realize any fully-digital beamformer by using a hybrid one. The literature on hybrid beamforming schemes is relatively vast. In [8] and [9], a point-to-point multiple-input multiple-output (MIMO) system is considered while the downlink of a multi-user setting is investigated in [10] using single-antenna receivers and a single-stream transmitter with a RF chain per user. In [11], the authors consider the more realistic case of imperfect channel state information due to the limited feedback of the return channel. All the aforementioned works are mainly focused on increasing the system spectral efficiency. There exist also some literature looking at reducing the power consumption.

Examples towards this direction can be found in [12] and [13]. In particular, [12] proposes the use of low-cost switches for implementing antenna selection schemes whereas [13] provides algorithms for selecting a subset of antennas. In [14], different hybrid architectures are compared in terms of both spectral and energy efficiency (EE), defined as the ratio between throughput and power consumption. Switching-based solutions are found to performs poorly compared to both fully-digital and hybrid schemes.

In addition to mobile communications, the main use cases of mmWave communications are wireless local area networks (WLANs) based on the IEEE 802.11ad standard as well as wireless backhaul in the unlicensed 60 Ghz band as a cost-efficient alternative to wired solutions. Wireless backhaul at mmWave bands is considered in [5], wherein the design of beam alignment techniques is investigated for a single-cell point-to-point network using an analog-only transceiver. Along this line of research, this work focuses on the downlink of a single-cell network in which a given number of multiple small-cell BSs exchange data with a macro BS through wireless backhaul, using a low-cost hybrid transceiver architecture [10], [11]. Our goal is to find respectively the optimal number $N$ and $M$ of antennas at the BS and each small-cell BS in order to maximize the EE. To this end, we first model the consumed power of a hybrid transceiver architecture at mmWave and then derive closed-form EE-optimal values for $M$ and $N$. These expressions provide valuable design insights into the interplay between system parameters and different components of the consumed power model. This work is inspired to the framework developed in [15], which however deals with the EE of massive MIMO networks and thus does not fit networks operating at mmWave frequencies.

The remainder of this work is organized as follows. Next section introduces the system model under a LoS channel propagation model and formulates the EE maximization problem. Section III develops the power consumption model of the hybrid transceiver network as a function of different system parameters. The EE-optimal number of antennas are computed in Section IV. Numerical results are given in Section V to validate the theoretical analysis. The numerical results are then extended to a more realistic clustered mmWave channel model in Section VI. Conclusions are drawn in Section VII.

## II. NETWORK MODEL AND PROBLEM STATEMENT

### A. Network model

We consider a two-tier network, which comprises a macro BS equipped with $N$ antennas and an overlaid tier of $K$ small-cell BSs (selected from a larger set) endowed with $M$ antennas and using a mmWave wireless backhaul link over a bandwidth $B$. We assume that the small-cell BSs are deployed so as to be in visibility with the macro BS. Due to the high absorption of scattered rays and the use of large antenna arrays (that create very narrow beam) at mmWave bands, a LoS model can be reasonably adopted for the propagation channel of each transmission link.[1] In these circumstances, the channel matrix $\mathbf{H}_k \in \mathbb{C}^{N \times M}$ between the BS and small-cell BS $k$ can be modeled as:

$$\mathbf{H}_k = \sqrt{\alpha_k}\mathbf{a}_N(\phi_k)\mathbf{a}_M^{\mathsf{H}}(\theta_k) \quad (1)$$

where $\mathbf{a}_N \in \mathbb{C}^{N \times 1}$ and $\mathbf{a}_M \in \mathbb{C}^{M \times 1}$ account, respectively, for the array manifolds of the BS and small-cell BSs with $\phi_k$ and $\theta_k$ being the angles of departure and arrival of the LoS link $k$. The parameter $\alpha_k$ describes the macroscopic pathloss and is computed as $\alpha_k = 10^{-l_{k,\mathrm{dB}}/10}$ with [5]

$$l_{k,\mathrm{dB}} = 32.5 + 20\log_{10} f_\mathrm{c} + 10\log_{10}(d_k)^\beta + Ad_k + \xi \quad (2)$$

where $f_\mathrm{c}$ [GHz] is the carrier frequency, $\beta$ is the pathloss exponent, $d_k$ [km] denotes the distance between the BS and small-cell BS $k$, $A$ accounts for the oxygen absorption and rainfall effect whereas $\xi \sim \mathcal{CN}(0, \sigma_\xi^2)$ is the shadowing being complex circularly symmetric Gaussian with variance $\sigma_\xi^2$.

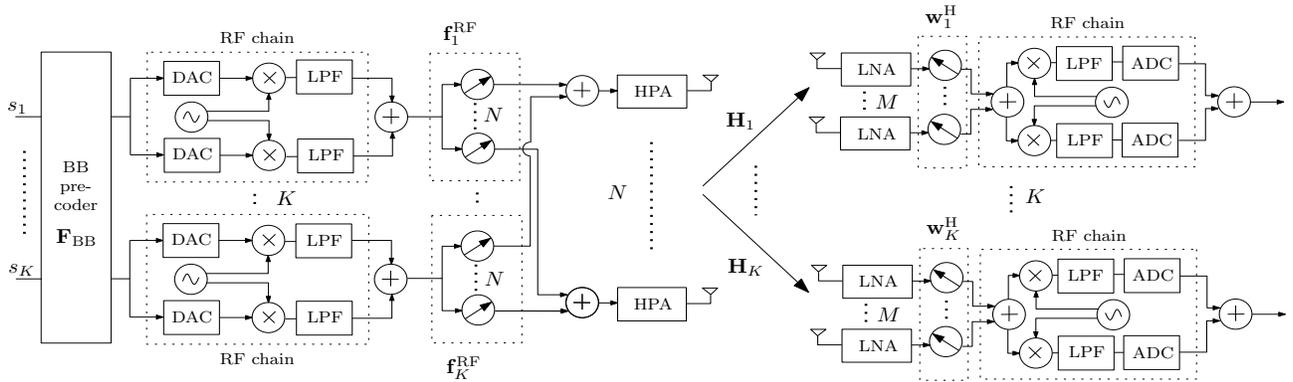

Fig. 1: Transceiver chain architecture.

Channel acquisition at mmWave bands is generally a challenging task due to the large number of antennas and the high bandwidth. However, if an uniform linear array (ULA) is adopted at both sides, the channel acquisition problem simply reduces to estimating the sets of directions $\{\theta_k, \phi_k\}$ and pathlosses $\{\alpha_k\}$ cutting down the number of unknowns from $(NM)K$ to $3K$. If mmWave communications are used for wireless backhaul, then channel estimation simplifies further due to the absence of mobility and the favorable deployment of the macro BS and small-cell BSs. In these circumstances, perfect channel state information seems to be a reasonable assumption (e.g., [5] and [17]). Based on this observation, in this work we assume perfect knowledge of $\{\theta_k, \phi_k, \alpha_k\}$. To limit the implementation costs [11], we assume that a two-stage linear hybrid precoding scheme is employed at the BS and that a RF linear combiner is used at each small-cell BS (see Fig. 1). In particular, the BS employs a baseband precoder $\mathbf{F}_{\mathrm{BB}} = [\mathbf{f}_1^{\mathrm{BB}}, \cdots, \mathbf{f}_K^{\mathrm{BB}}] \in \mathbb{C}^{R \times K}$ followed by a RF precoder $\mathbf{F}_{\mathrm{RF}} = [\mathbf{f}_1^{\mathrm{RF}}, \cdots, \mathbf{f}_R^{\mathrm{RF}}] \in \mathbb{C}^{N \times R}$ with $K \le R \le N$ being the number of RF chains. The transmitted vector $\mathbf{x} \in \mathbb{C}^N$ is thus given by $\mathbf{x} = \mathbf{F}_{\mathrm{RF}}\mathbf{F}_{\mathrm{BB}}\mathbf{s}$ where $\mathbf{s} \in \mathbb{C}^K$ is the data vector such that $\mathbb{E}\{\mathbf{s}\mathbf{s}^{\mathsf{H}}\} = P/N \mathbf{I}_K$ with $P$ being the transmitted power. Hereafter, we assume that $R = K$, i.e. one stream per small-cell BS is allocated.

[1]Observe that the LoS condition is also valid in highly dense mmWave networks, where having links in visibility is more likely to happen [16].

At small-cell BS $k$, the received signal is linearly processed through the RF combiner $\mathbf{w}_k$ to obtain:

$$y_k = \mathbf{w}_k^{\mathsf{H}}\mathbf{H}_k^{\mathsf{H}}\mathbf{x} + \mathbf{w}_k^{\mathsf{H}}\mathbf{n}_k \quad (3)$$

where $\mathbf{n}_k \sim \mathcal{CN}(\mathbf{0}, \sigma^2\mathbf{I}_M)$ is the thermal noise with $\sigma^2 = BN_0 N_\mathrm{F}$ [W] while $N_0$ [W/Hz] and $N_\mathrm{F}$ being the noise power spectral density and noise figure, respectively. The RF combiners $\{\mathbf{w}_k\}$ and precoders $\{\mathbf{f}_k^{\mathrm{RF}}\}$ are implemented using analog phase shifters. Under the assumption of perfect knowledge of $\{\theta_k, \phi_k\}$, we have that $\mathbf{w}_k = \mathbf{a}_M(\theta_k)$ and $\mathbf{f}_k^{\mathrm{RF}} = \mathbf{a}_N(\phi_k)$. Therefore, $y_k$ reduces to:

$$y_k = (MN)\bar{\mathbf{h}}_k^{\mathsf{H}}\mathbf{F}_{\mathrm{BB}}\mathbf{s} + \mathbf{a}_M^{\mathsf{H}}(\theta_k)\mathbf{n}_k \quad (4)$$

where $\bar{\mathbf{h}}_k^{\mathsf{H}} = \frac{\sqrt{\alpha_k}}{N}\mathbf{a}_N^{\mathsf{H}}(\phi_k)\mathbf{F}_{\mathrm{RF}}$ is the *effective* channel seen from small-cell BS $k$ after receive combining. The BB precoder $\mathbf{F}_{\mathrm{BB}}$ is designed according to a zero-forcing (ZF) criterion so as to completely remove the interference among small-cell BSs [11]. This leads to $\mathbf{F}_{\mathrm{BB}} = (\bar{\mathbf{H}}^{\mathsf{H}})^{-1}$ where $\bar{\mathbf{H}}^{\mathsf{H}} = [\bar{\mathbf{h}}_1, \ldots, \bar{\mathbf{h}}_K]^{\mathsf{H}} = \frac{1}{N}\mathbf{D}^{1/2}(\mathbf{F}_{\mathrm{RF}}^{\mathsf{H}}\mathbf{F}_{\mathrm{RF}})$ with $\mathbf{D} = \mathrm{diag}(\alpha_1, \ldots, \alpha_K)$. Plugging $\mathbf{F}_{\mathrm{BB}} = (\bar{\mathbf{H}}^{\mathsf{H}})^{-1}$ into (4) yields

$$y_k = (MN)s_k + \mathbf{a}_M^{\mathsf{H}}(\theta_k)\mathbf{n}_k. \quad (5)$$

Note that the inverse of $\bar{\mathbf{H}}^{\mathsf{H}}$ exists as long as $\phi_l - \phi_k \ne 0$ for $k, l = 1, \ldots, K$, which always occurs in practice if the served small-cell BSs are properly selected.

### B. Problem statement

The aim of this work is to compute the values of $(N, M)$ that, for a given number $K$ of small-cell BSs, maximize the

EE of the network given by:

$$\text{EE} = \frac{\text{Throughput}}{\text{Consumed Power}} \quad \text{[bit/Joule]} \quad (6)$$

which stands for the number of bits that can be reliably transmitted per unit of energy. From (5), the throughput of the considered network is easily found as:

$$\text{Throughput} = BK \log_2(1 + MN\gamma) \quad \text{[bit/s]} \quad (7)$$

with $\gamma = P/\sigma^2$. Observe that we have neglected the pre-log factor that should take into account the signaling overhead for channel estimation, due to the stationarity of the investigated network [15]. The consumed power is computed as [15]

$$\text{Consumed Power} = \eta^{-1}P_\mathbf{x} + P_\text{CP} \quad \text{[W]} \quad (8)$$

where $P_\mathbf{x}$ is the transmit power, $\eta \leq 1$ is the power amplifier (PA) efficiency and $P_\text{CP}$ accounts for the power consumed by the circuitry.

## III. POWER CONSUMPTION MODEL

A reasonable circuit power consumption model for a generic BS in a cellular network is as follows [15]

$$P_\text{CP} = P_\text{FIX} + P_\text{TC} + P_\text{LP} + P_\text{CE} + P_\text{C/BH} \quad (9)$$

where $P_\text{FIX}$ accounts for the fixed power consumption of the system, $P_\text{TC}$ of the transceiver chain (at both BS and small-cell sides), $P_\text{CE}$ of the channel estimation process, $P_\text{LP}$ of the linear processing, $P_\text{C/BH}$ of the coding at BS and of the load-dependent backhauling cost. Next all the above terms will be explicated as a function of all the system parameters in Table I taken for a reference carrier frequency of $f_\text{c} = 60$ Ghz.

### A. Transmitted power

The average transmit power is given by $P_\mathbf{x} = \mathbb{E}\{\|\mathbf{x}\|_2^2\}$ where the expectation is taken with respect to the set of distances $\mathbf{d} = [d_1, \cdots, d_K]$ and AoDs $\boldsymbol{\phi} = [\phi_1, \cdots, \phi_K]$, and thus, can be computed as

$$\begin{aligned}
P_\mathbf{x} &= \text{tr}\left(\mathbb{E}\{\mathbf{ss}^\mathsf{H}\}\mathbb{E}\{\mathbf{F}_\text{BB}^\mathsf{H}\mathbf{F}_\text{RF}^\mathsf{H}\mathbf{F}_\text{RF}\mathbf{F}_\text{BB}\}\right) \\
&= \frac{P}{N}\text{tr}\left(\mathbb{E}\{(\bar{\mathbf{H}}^\mathsf{H})^{-1}(N\mathbf{D}^{-1/2}\bar{\mathbf{H}}^\mathsf{H})(\bar{\mathbf{H}}^\mathsf{H})^{-1}\}\right) \\
&= NP\,\text{tr}\left(\mathbb{E}\{\mathbf{D}^{-1}(\mathbf{F}_\text{RF}^\mathsf{H}\mathbf{F}_\text{RF})^{-1}\}\right) \\
&= NP \sum_{k=1}^{K}\mathbb{E}_\mathbf{d}\{\alpha_k^{-1}\}\mathbb{E}_\phi\left\{\left[(\mathbf{F}_\text{RF}^\mathsf{H}\mathbf{F}_\text{RF})^{-1}\right]_{k,k}\right\} \\
&\stackrel{(a)}{=} NPK\bar{\alpha}\mathbb{E}_\phi\left\{\left[\mathbf{P}^{-1}\right]_{k,k}\right\} \quad (10)
\end{aligned}$$

where (a) follows from assuming that any small-cell location is drawn from the same spatial distribution such that $\bar{\alpha} = \mathbb{E}_\mathbf{d}\{\alpha_k^{-1}\}$. Also, we have defined for notational simplicity $\mathbf{P} = \mathbf{F}_\text{RF}^\mathsf{H}\mathbf{F}_\text{RF} \in \mathbb{C}^{K \times K}$. A possible way to deal with the computation of $\mathbb{E}_\phi\{[\mathbf{P}^{-1}]_{k,k}\}$ is to make use of the Kantorovic inequality [11], which reads (exploiting the fact that $[\mathbf{P}]_{k,k} = 1$)

$$\begin{aligned}
\left[\mathbf{P}^{-1}\right]_{k,k} &\leq \frac{1}{4[\mathbf{P}]_{k,k}}\left(\kappa(\mathbf{P}) + \kappa(\mathbf{P})^{-1} + 2\right) \\
&= \frac{1}{4N}\left(\kappa(\mathbf{P}) + \kappa(\mathbf{P})^{-1} + 2\right) \quad (11)
\end{aligned}$$

where $\kappa(\mathbf{P}) = \kappa^2(\mathbf{F}_\text{RF})$ and $\kappa(\mathbf{F}_\text{RF}) = \|\mathbf{F}_\text{RF}\|\|\mathbf{F}_\text{RF}^\dagger\|$ stands for the 2-norm condition number of the Vandermonde matrix with entries $[\mathbf{F}_\text{RF}]_{n,k} = z_k^n$ for $n = 0, \cdots, N-1$ and *nodes* $\{z_k\}_{k=1}^K = e^{j\pi\sin(\phi_k)}$ for normalized antenna spacing $\Delta/(2\pi f_\text{c}) = 1/2$. Computing $\kappa(\mathbf{F}_\text{RF})$ is a challenge, especially because the analysis must be valid for the entire range of antennas. Vandermonde matrices with positive real nodes $z_k \in \mathbb{R}_+$ are well-known to be ill-conditioned [23] - the condition number grows at least exponentially with the number of nodes $K$. However, if the nodes are complex-valued, it is possible to lower this growth to polynomial [24] and even achieve perfect conditioning choosing the nodes to be roots of unity [25]. In [26], the authors generalize this result to nodes that are close enough to the unit circle (not necessarily on the unit circle) and not so close to each other, while having $N$ large enough. In particular, it turns out that if $|z_k| = 1$ and $N > 2\frac{K-1}{\delta}$ then [26]

$$1 \leq \kappa(\mathbf{F}_\text{RF}) \leq \frac{1 + \frac{2}{\delta}\frac{K-1}{N}}{1 - \frac{2}{\delta}\frac{K-1}{N}} \quad (12)$$

with $\delta = \min_{j \neq k}|z_j - z_k|$ accounting for the worst-case node separation. Thus, in order for the Vandermonde matrix $\mathbf{F}_\text{RF}$ to be nearly perfect conditioned we better impose

$$N \gg 2\frac{K-1}{\delta}. \quad (13)$$

To get some insight into how much large $N$ should be, we consider a uniformly spaced small-cell deployment on the right side quadrants and evaluate $\delta$. If the small-cell BSs are such that $\{\phi_k\}_{k=1}^K = -\frac{\pi}{K}\lfloor\frac{K}{2}\rfloor + \frac{\pi}{K}$ then

$$\begin{aligned}
\delta &= |z_{\lfloor\frac{K}{2}\rfloor} - z_{\lfloor\frac{K}{2}\rfloor-1}| = |1 - e^{j\pi\sin(\phi_1)}| \\
&= 2\left|\sin\left(\frac{\pi}{2}\sin\left(\frac{\pi}{K}\right)\right)\right| \quad (14)
\end{aligned}$$

from which it follows that, when $K$ grows large, $\delta$ can be well-approximated with $\pi^2/K$ (using first order Taylor expansion). Plugging $\pi^2/K$ into (13) leads to $N \gg 2K(K-1)/\pi^2$. This means that, for $K$ sufficiently large, the value of $N$ for achieving good conditioning for $\mathbf{F}_\text{RF}$ is given by

$$N \geq \frac{2\lambda}{\pi^2}K^2 = \mu_K \quad (15)$$

with $\lambda \geq 1$ being a design parameter. Under this condition, by using (11) and (15) into (10) we have that $P_\mathbf{x}$ can be reasonably approximated as

$$P_\mathbf{x} = PK\bar{\alpha} \quad \text{for} \quad N \geq \mu_K. \quad (16)$$

Fig. 2 illustrates $\kappa(\mathbf{F}_\text{RF})$ as a function of $N$ for different values of $K$ and uniformly spaced nodes, i.e. $\{\phi_k\}_{k=1}^K = -\frac{\pi}{K}\lfloor\frac{K}{2}\rfloor +$ $\frac{\pi}{K}$. As seen, $\kappa(\mathbf{F}_{\text{RF}})$ tends to unity when $N$ grows for any $K$. Also, it can be seen numerically that $\lambda = 1$ is already enough to satisfy condition (15) when the nodes (small-cell BSs) are properly selected. A similar behavior is observed for AoDs uniformly distributed $\phi_k \in \mathcal{U}[-\pi/2, \pi/2]$ [27].

TABLE I: Network and system parameters at 60GHz.

| Parameter | Description | Value | Parameter | Description | Value |
|---|---|---|---|---|---|
| $P_{\text{LNA}}$ | Power consumed by low noise amplifier [18] | 39 [mW] | $L_{\text{BH}}$ | Power used by backhauling per bit/s [15] | 250 [mW/Gbit/s] |
| $P_{\text{HPA}}$ | Power consumed by the high-power-amplifier [18], [19] | 138 [mW] | $T_c$ | Coherence time [20] | 10 [s] |
| $P_{\text{DC}}$ | Power absorbed by the down conversion stage [21] | 47.3 [mW] | $\Delta$ | Normalised antenna separation | 0.5 |
| $P_{\text{UC}}$ | Power absorbed by the up conversion stage [19] | 49 [mW] | $\sigma_\xi^2$ | Shadowing variance [5] | 8 [dB] |
| $P_{\text{ADC}}$ | Power needed to run the analog-to-digital converter [12] | 200 [mW] | $A$ | Oxigen and rainfall absorption [5] | 25 [dB] |
| $P_{\text{DAC}}$ | Power needed to run the digital-to-analog converter [22] | 110 [mW] | $\kappa$ | Path-loss exponent [5] | 2.2 |
| $P_C$ | Power consumed by the combiner [18] | 19.5 [mW] | $N_0$ | Noise power spectral density [5] | -174 [dBm] |
| $P_{\text{PS}}$ | Power required to commute phase shifter [18] | 30 [mW] | $d$ | Distance BS to small-cell BSs [5] | 150 [m] |
| $L_{\text{BS}}$ | Computational efficiency at the BS [15] | 20 [Gflops/W] | $N_F$ | Noise figure [5] | 6 [dB] |
| $L_{\text{SC}}$ | Computational efficiency at the small-cell BSs [15] | 5 [Gflops/W] | $B$ | Transmission bandwidth [5] | 2 [Ghz] |
| $L_C$ | Power consumed performing coding per bit/s [15] | 100 [mW/Gbit/s] | $f_c$ | Carrier frequency [5] | 60 [Ghz] |
| $L_D$ | Power consumed performing decoding per bit/s [15] | 800 [mW/Gbit/s] | $\eta$ | High power amplifier efficiency [15] | 0.375 |

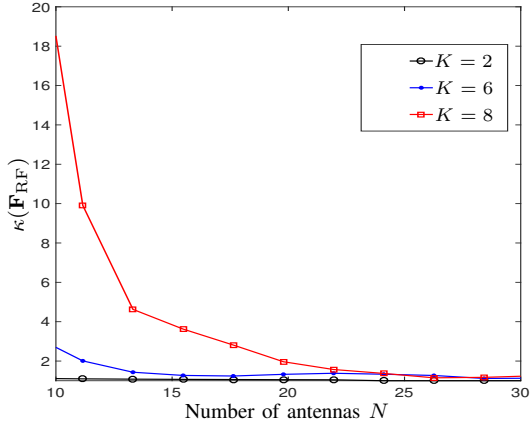

Fig. 2: Condition number of $\mathbf{F}_{\text{RF}}(\boldsymbol{\phi})$ versus $N$ for $K = 2, 6$ and $8$.

### B. Transceiver Chain

The transceiver architecture of the investigated network is sketched in Fig.1. We assume that both the BS and the set of small-cell BSs make use of (at least) 5-bit passive phase shifters (PSs) that emulate the arbitrary angles matching at RF [12]. Each small-cell BS consists of a single RF chain connected through a combiner to $M$ parallel front-end (FE) receivers, one for each receive antenna. Each FE receiver is composed of a low-noise amplifier (LNA) followed by a phase shifter, while an RF chain hosts a couple (I/Q) of analog-to-digital converters (ADCs), and a down conversion stage that includes a mixer, a voltage controlled oscillator and a baseband buffer [21]. Therefore, the power consumption of the transceiver chain at each small-cell BS can be computed as

$$P_{\text{TC}}^{\text{SC}} = \underbrace{M(P_{\text{LNA}} + P_{\text{PS}})}_{\text{Front-end}} + \underbrace{P_{\text{DC}} + P_{\text{ADC}} + P_C}_{\text{RF chain}} \quad (17)$$

where $P_{\text{LNA}}$ accounts for the power consumption of each LNA, $P_{\text{PS}}$ of each PS, $P_{\text{DC}}$ of the down-conversion, $P_{\text{ADC}}$ of the ADC and $P_C$ of the combiner.

On the other hand, the BS transceiver consists of $K$ RF chains each one fetching a rake of $N$ PSs that drive the phases of $N$ antennas, each one with a high power amplifier (HPA). Each RF chain has a pair (I/Q) of digital-to-analog converters (DACs) plus a combiner as well as an up-conversion stage including filtering and amplifying. Therefore, we have that

$$P_{\text{TC}}^{\text{BS}} = \underbrace{N(KP_{\text{PS}} + P_{\text{HPA}} + P_C)}_{\text{Front-end}} + \underbrace{K(P_{\text{UC}} + P_{\text{DAC}})}_{\text{RF chain}} \quad (18)$$

Therefore, the total amount of consumed power in the transceiver chain is

$$P_{\text{TC}} = P_{\text{TC}}^{\text{BS}} + KP_{\text{TC}}^{\text{SC}} = p_{\text{RF}} + p_{\text{FE}}^{\text{SC}}M + p_{\text{FE}}^{\text{BS}}N \quad (19)$$

where $p_{\text{RF}} = K(P_{\text{DC}} + P_{\text{ADC}} + P_C + P_{\text{UC}} + P_{\text{DAC}})$ accounts for the power consumption of the RF chain at both sides, whereas $p_{\text{FE}}^{\text{SC}} = K(P_{\text{LNA}} + P_{\text{PS}})$ and $p_{\text{FE}}^{\text{BS}} = KP_{\text{PS}} + P_{\text{HPA}} + P_C$ of the FEs at the small-cells BS and BS, respectively.

### C. Linear Processing

The power consumed by linear processing accounts for all the operations performed in the digital domain at the macro BS. This be quantified as

$$P_{\text{LP}} = \underbrace{P_{\text{LP-T}}}_{\text{Transmission}} + \underbrace{P_{\text{LP-P}}}_{\text{Precoder computation}} \quad (20)$$

where $P_{\text{LP-T}}$ accounts for the total power consumed by downlink transmission of payload samples whereas $P_{\text{LP-P}}$ is the power required for the computation of $\mathbf{F}_{\text{BB}}$. Due to the stationarity of the investigated network, the latter can be neglected since it is computed once for all. This amounts to saying that $P_{\text{LP-P}} = 0$. The computation of $\mathbf{F}_{\text{BB}}$s requires a total of $K(2K-1)$ complex operations per sample. Denoting by $L_{\text{BS}}$ the computational efficiency of the BS [flops/W], we have that

$$P_{\text{LP}} = B\frac{K(2K-1)}{L_{\text{BS}}}. \quad (21)$$

### D. Coding/Decoding and Backhauling

Load-dependent power costs are given by coding/decoding and backhauling. In the downlink, the BS applies channel coding and modulation to $K$ sequences of information symbols and each small-cell BS applies some suboptimal fixed-complexity algorithm for decoding its own sequence. The opposite is done in the uplink. The power consumption accounting for these processes is proportional to the number of bits. The backhaul is used to transfer uplink/downlink data between the BS and the core network. The power consumption of the backhaul is commonly modeled as the sum of two parts: one load-independent (included in the fix power consumption) and one load-dependent (proportional to the throughput). Therefore, the power consumption for coding/decoding and backhauling processes can be computed as

$$P_{\text{C/BH}} = L_{\text{B}} B K \log_2(1 + MN\gamma) \quad (22)$$

where $L_{\text{B}} = L_{\text{C/D}} + L_{\text{BH}}$ with $L_{\text{C/D}}$ and $L_{\text{BH}}$ being the operational costs for coding/decoding and backhauling, respectively.

## IV. EE OPTIMIZATION

Plugging (7)-(9) and (16)-(22) into (6), the EE optimization problem can thus be formulated as

$$\underset{(M,N)\in\mathbb{Z}_{++}}{\arg\max}\ \text{EE}(M,N,K) \quad \text{s.t.} \quad N \geq \mu_K \quad (23)$$

with

$$\text{EE} = \frac{BK\log_2(1+\gamma MN)}{\bar{P}_{\text{FIX}} + p_{\text{FE}}^{\text{SC}} M + p_{\text{FE}}^{\text{BS}} N} \quad (24)$$

and

$$\bar{P}_{\text{FIX}} = p_{\text{RF}} + P_{\text{FIX}} + P_{\mathbf{x}}\eta^{-1} + P_{\text{LP}}. \quad (25)$$

In the following, we aim at solving (23) for fixed system parameters as given in Table I. In doing so, we first derive a closed-form expression for the EE-optimal value of both $M$ and $N$ when the other one is fixed. This does not only bring indispensable insights into the interplay between $(M,N)$ and the system parameters, but provides the means to solve the problem by a sequential optimization algorithm.

### A. Optimum number of small-cell BS antennas

We begin by deriving the optimal number of small-cell BS antennas $M$ while $N$ is fixed. Applying [15, Lemma 3], it readily follows that:

**Lemma 1.** *Assume $N$ is given, then the optimal $M$ can be computed as $M^\star = \lfloor x^\star \rceil$ with*

$$x^\star = \frac{e^{\mathcal{W}\left(\frac{\gamma}{e}\frac{c_1}{c_2} - \frac{1}{e}\right)+1} - 1}{\gamma N} \quad (26)$$

*and $c_1 = N\left(\bar{P}_{\text{FIX}} + p_{\text{FE}}^{\text{BS}} N\right)$, $c_2 = p_{\text{FE}}^{\text{SC}}$ and $\lfloor \cdot \rceil$ as the nearest integer projector.*

The above result provides explicit guidelines on how to select $M$ in a hybrid mmWave system for maximal EE. Notice that the term $c_1$ depends, through $\bar{P}_{\text{FIX}}$, on $p_{\text{RF}}$, which accounts for the RF chain power consumption of the transceiver architecture, and also on the front-end power consumption $p_{\text{FE}}^{\text{BS}}$ at the BS. Using the typical values of Table I, it turns out that $c_1$ is on the order of hundreds of Watt for a relatively small number of antennas $N$. Larger values are obtained if $N$ increases. On the other hand, $c_2$ does not depend on $N$ and takes values in the range of Watt, since it depends only on the power consumed by the small-cell BSs for the front-end. Therefore, we can reasonably assume that, for typical values of system parameters, $c_1/c_2 \gg 1$ such that $e^{\mathcal{W}(r)+1}$ can be approximated[2] with $r$ and $x^\star$ reduces to

$$x^\star \approx \frac{1}{N}\frac{1}{e}\frac{c_1}{c_2} = \frac{\bar{P}_{\text{FIX}} + p_{\text{FE}}^{\text{BS}} N}{e\, p_{\text{FE}}^{\text{SC}}}. \quad (27)$$

Using the above result and the power consumption expressions provided in Section III, the following corollary is found:

**Corollary 1.** *If $N$ and $K$ grow large, then $M^\star$ increases monotonically as:*

$$M^\star \approx \left\lfloor \xi + \frac{p_{\text{FE}}^{\text{BS}}}{p_{\text{FE}}^{\text{SC}}} N \right\rceil \quad (28)$$

*with $\xi = \left(p_{\text{RF}} + P_{\text{FIX}} + \frac{2B}{L_{\text{BS}}}K^2\right)/p_{\text{FE}}^{\text{SC}}$, $p_{\text{FE}}^{\text{SC}}$ and $p_{\text{FE}}^{\text{BS}}$ as in (17) and (18), respectively.*

From the above corollary, it follows that $M^\star$ is monotonically increasing with $P_{\text{FIX}}$ as well as with $K$ and $N$. Using the values reported in Table I, it turns out that $p_{\text{FE}}^{\text{BS}}/p_{\text{FE}}^{\text{SC}} < 1$, meaning that $M^\star$ grows at a slower pace than $N$. Also, the term $\xi$ indicates that $M^\star$ increases linearly with $p_{\text{RF}}$, i.e., the power consumed by the FE at both the BS and small-cell BSs.

### B. Optimum number of BS antennas

We now look for the value of $N$ that maximizes the EE in (23). Still, by using [15, Lemma 3] and exploiting the pseudo concavity of the objective function, the following result is obtained:

**Lemma 2.** *Assume $M$ is given, then the optimal $N$ is given by $N^\star = \lfloor z^\star \rceil$ with*

$$z^\star = \max\left\{ \frac{e^{\mathcal{W}\left(\frac{\gamma}{e}\frac{d_1}{d_2} - \frac{1}{e}\right)+1} - 1}{\gamma M}, \mu_K \right\} \quad (29)$$

*and $d_1 = M\left(\bar{P}_{\text{FIX}} + p_{\text{FE}}^{\text{SC}} M\right)$, $d_2 = p_{\text{FE}}^{\text{BS}}$, $\mu_K$ as in (15) and $\lfloor \cdot \rceil$ as the nearest integer projector.*

As for $M$, we have that $z^\star$ can be reasonably approximated as $z^\star \approx \frac{1}{N}\frac{1}{e}\frac{d_1}{d_2}$ from which it follows that:

**Corollary 2.** *If $M$ and $K$ grow large, then $N^\star$ increases monotonically as:*

$$N^\star \approx \left\lfloor \max\left\{ \xi + \frac{p_{\text{FE}}^{\text{SC}}}{p_{\text{FE}}^{\text{BS}}} M, \mu_K \right\} \right\rceil \quad (30)$$

In agreement with the results of Corollary 1, we have that $N^\star$ grows at faster pace than $M$ since $p_{\text{FE}}^{\text{SC}}/p_{\text{FE}}^{\text{BS}} > 1$ as it follows using the values of Table I. Therefore, using larger arrays at the BS rather than at small-cell BSs seems to be a more natural choice for maximal EE.

---
[2]The interested reader is referred to [28] for further details on the inequalities and approximations involving the Lambert function.

## C. Sequential Optimization of $M, N$

Using Lemmas 1 and 2, a sequential optimization algorithm to solve (23) operates as follows:

1) Optimize $M$ for a fixed $N$ using Lemma 1;
2) Optimize $N$ for a fixed $M$ using Lemma 2;
3) Repeat 1)–2) until convergence is achieved.

This algorithm converges since the EE is a non-decreasing monotone function of $(M, N)$ and bounded above. The monotonicity is ensured by the pseudo concavity of (24). Indeed, the numerator is non-negative, differentiable, and concave, while the denominator is differentiable and affine, and so convex.

## V. NUMERICAL RESULTS

Numerical results are now used to validate the analysis. We consider a single-cell scenario as described in Section II with a macro BS, operating at $f_c = 60$ GHz over a bandwidth of $B = 2$ GHz placed at the center of the cell and serving simultaneously $K$ small-cell BSs, with a distance $d = 150$ m from the BS. To avoid ambiguity in the spatial domain, the small cells are angularly displaced on the right half-space centered on the BS. The channel parameters and all of the terms introduced in Section III are listed in Table I. To make the numerical results as realistic as possible, the same fabrication technology (65nm CMOS[3]) is used for the circuit parameters (e.g. [18] and [12]), while the linear processing and the traffic-dependent parameters are from [15]. The channel model parameters are taken from [30] and [5]. Results are obtained for a signal-to-noise ratio of $\gamma = 0$ dB.

Fig. 3a shows the EE as a function of $M$ and $N$ when $K = 10$. We see that there is a global maximizer for $(M^\star, N^\star) = (19, 32)$ to which corresponds an $EE^\star = 620$ Mbit/Joule and a throughput of $18.4$ Gbit/s per small-cell BS. The total power consumed by circuitry is approximately $P_{\mathrm{CP}} = 290$ W. The sequential optimization algorithm described in Section IV converges after a few iterations to the global optimizer validating (16). As seen, the optimal configuration is characterized by a relatively small $N^\star = 30$, which is slightly larger than the number of served small cells, i.e. $K = 10$. In other words, the output of the optimization problem suggests to use a number of BS antennas that is on the same order of magnitude of $K$. This is in contrast to what it is usually required in mmWave communications for maximal spectral efficiency, namely, a large antenna array at both sides of the link to cope with the severe propagation conditions. To be energy-efficient, the so-called doubly massive MIMO paradigm[4] requires either better beamforming schemes (increasing the throughput) or more power efficient electronic devices (reducing the power consumption). This latter case is investigated in Fig. 3b in which the power consumed by front-end devices is decreased by an order of magnitude, both at the BS ($p_{\mathrm{FE}}^{\mathrm{BS}}$) and at the small-cell BSs ($p_{\mathrm{FE}}^{\mathrm{SC}}$). We see that in this case a doubly massive

[3]CMOS technology promises higher levels of integration and reduced cost with respect to other solutions on the market such as GaAs and InP [29].

[4]In literature doubly massive MIMO is referred to a system equipped with very large antenna arrays at both transmitter and receiver.

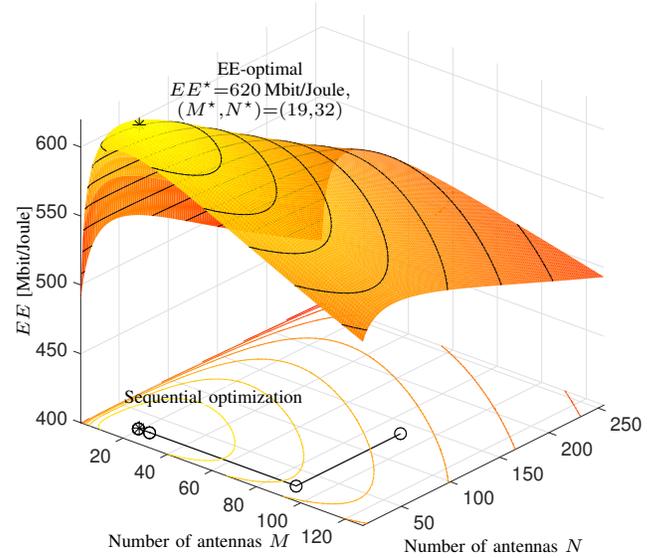

(a) $p_{\mathrm{FE}}^{\mathrm{BS}}$ and $p_{\mathrm{FE}}^{\mathrm{SC}}$ as in Table I

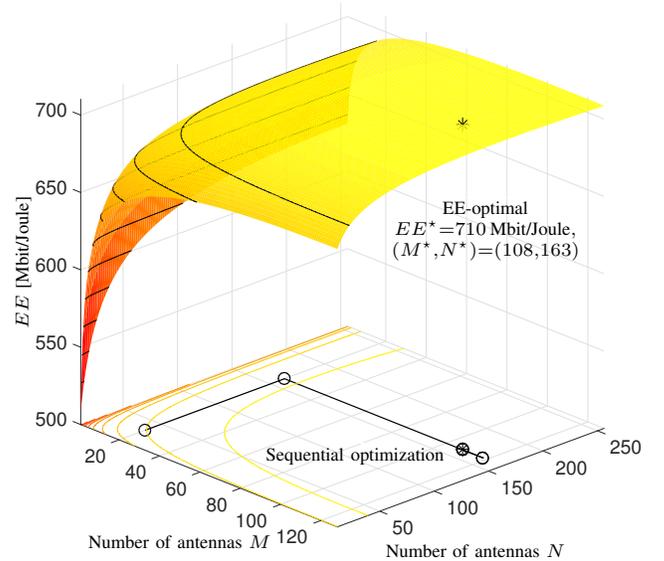

(b) $p_{\mathrm{FE}}^{\mathrm{BS}}$ and $p_{\mathrm{FE}}^{\mathrm{SC}}$ as in Table I scaled by a factor $10\times$

Fig. 3: Energy Efficiency [Mbit/Joule] for different combination of $M$ and $N$ (with $K = 10$).

MIMO setup with $(M^\star, N^\star) = (108, 163)$ naturally arises at the EE-optimal. The throughput is also increased by a factor $1.5\times$ with respect to the EE-optimal in Fig. 3a. Based on the above results, it follows that, to improve the EE and throughput of mmWave communications, the hardware components (such as PSs, LNAs and HPAs) have to be more efficient than todays.

## VI. EXTENSION TO NLOS CHANNELS

In this section, we investigate to what extent the major conclusions can be extended to a NLoS scenario.

## A. Network model

We adopt a time-invariant clustered channel model composed of a LoS path and $N_{\text{cl}}$ scattering clusters, each one contributing with $N_r$ rays accounting for the NLoS component. This leads to the following channel matrix $\mathbf{H}_k \in \mathbb{C}^{N \times M}$ between the BS and small-cell BS $k$:

$$\mathbf{H}_k = \frac{1}{\sqrt{N_{\text{cl}} N_r}} \sum_{i=1}^{N_{\text{cl}}} \sum_{j=1}^{N_r} \sqrt{\alpha_{i,j,k}} \mathbf{a}_N(\phi_{i,j,k}) \mathbf{a}_M^{\mathsf{H}}(\theta_{i,j,k}) + I_{\text{LoS}}(d_k) \sqrt{\alpha_k} \mathbf{a}_N(\phi_{k,\text{LoS}}) \mathbf{a}_M^{\mathsf{H}}(\theta_{k,\text{LoS}}) \quad (31)$$

where $\phi_{i,k}$ and $\theta_{i,k}$ are the mean AoD and AoA of each link between BS and the $i$-th scatterer. The angle spread within each cluster is also taken into account by using Laplacian distribution, $\phi_{i,j,k} \sim \mathcal{L}(\phi_{i,k}, \mu_{i,k})$ and $\theta_{i,j,k} \sim \mathcal{L}(\theta_{i,k}, \mu_{i,k})$. The parameter $\alpha_{i,j,k}$ includes both the small-scale and the large-scale fading effect and is computed as $\alpha_{i,j,k} = \tilde{\alpha}_{i,j,k} 10^{-l_{i,k,\text{dB}}/10}$ with $l_{i,k,\text{dB}}$ as in (2) and $\tilde{\alpha}_{i,j,k}$ accounting or the small-scale effects. The set of NLoS distances can be evaluated by geometrical considerations as

$$d_{i,k} = d_{i,k}^{\text{cl}} + \sqrt{(d_{i,k}^{\text{cl}} \sin \bar{\phi}_{i,k})^2 + (d_k - d_{i,k}^{\text{cl}} \cos \bar{\phi}_{i,k})^2} \quad (32)$$

where $d_{i,k}^{\text{cl}}$ and $d_k$ are the distances BS-cluster $i$ (when pointing small cell $k$) and BS-small cell $k$, respectively and $\bar{\phi}_{i,k} = \phi_{i,k} - \phi_{k,\text{LoS}}$, $\bar{\theta}_{i,k} = \theta_{i,k} - \theta_{k,\text{LoS}}$. Besides, in the LoS component, $I_{\text{LoS}} \sim \mathcal{B}(p(d_k))$ is a Bernoulli random variable indicating the presence or not of the LoS link[5]. Unlike the NLoS component, $\theta_{k,\text{LoS}}$ and $\phi_{k,\text{LoS}}$ are related as $\theta_{k,\text{LoS}} = \text{mod}(\pi + \phi_{k,\text{LoS}}, 2\pi)$. We refer to [30] and [3] for further details. Hereafter, to dimension the precoder/combiner we use the same eigenmode beamforming approach used in Section II, in the analog domain, along with a digital ZF precoder. In particular, let $\mathbf{H}_k^{\mathsf{H}} = \mathbf{U}_k \boldsymbol{\Sigma}_k \mathbf{V}_k^{\mathsf{H}}$ be the singular value decomposition (SVD) of $\mathbf{H}_k^{\mathsf{H}}$, the k-th user precoding and combining vectors, $\mathbf{f}_{\text{RF},k}$ and $\mathbf{w}_k$, are chosen as the columns of the matrices $\mathbf{V}_k$ and $\mathbf{U}_k$ corresponding to the largest eigenvalue of $\boldsymbol{\Sigma}_k$, i.e. $\mathbf{v}_{k,1}$ and $\mathbf{u}_{k,1}$. We then project the beamforming matrices $\mathbf{F}_{\text{RF}}$ and $\mathbf{W}$ onto the analog set $\mathcal{S}_{p,q} = \{\mathbf{X} \in \mathbb{C}^{p \times q} : |\mathbf{X}_{i,j}| = 1, (i,j) = \{p\} \times \{q\}\}$. This simply results in scaling each entry of those matrices by its magnitude [8]. The precoder $\mathbf{F}_{\text{BB}}$ is designed according to a ZF criterion to cope with the effective interference after analog precoding-combining.

## B. Numerical results

Fig. 4a shows numerically how the EE behaves as a function of $M$ and $N$ using the NLoS channel model described above. The optimal operating point is found at $(M^\star, N^\star) = (5, 30)$ to which corresponds an $EE = 709$ Mbit/Joule, an aggregate throughput and circuit power consumption are respectively 29.2 Gbit/s per small-cell BS and 412 W. The above network configuration is far from being considered as doubly-massive MIMO. This supports our conclusion that such systems, when used with hybrid architectures, are not optimal from an EE perspective. Fig. 4b illustrates the EE of a fully-digital system, which applies the ZF precoder entirely in the baseband, that is $\mathbf{F}_{\text{RF}} \mathbf{F}_{\text{BB}} = \mathbf{F} = \bar{\mathbf{H}}^\dagger$. In addition, to fairly compare the performance of the fully-digital to that of the hybrid scheme in Fig. 1, constant transmit power at BS is ensured, i.e. $\|\mathbf{x}\|_2^2 = \|\mathbf{s}\|_2^2$. The transmitted vector of symbols must be changed accordingly so as $\mathbf{s}' = (\mathbf{1}_M \otimes \mathbf{I}_K) \mathbf{s}$. At the small cell side, linear combining is performed by matching the most significant left eigenvector of the channel $\mathbf{w}_k = \mathbf{u}_{k,1}$ associated to the highest eigenvalue. Fig. 4b further validates the tendency encountered for the hybrid system, which is to avoid the use of large arrays at both network sides. Here, the EE-optimal point is at $(M^\star, N^\star) = (1, 24)$ achieving a throughput of 27.2 Gbit/s per small-cell BS with 381 W of consumed power. Although the precoders perform similarly, the hybrid solution leads to a smoother EE function that is preferable for its robustness to system changes. Moreover, Table II shows how much the circuit power terms contribute to the overall consumed power at the EE-optimal, both for the hybrid and fully-digital case. As we can see, in the hybrid case, the major contribution comes from the fixed power, while in the fully-digital one it comes from the power drawn by the FE chain at the BS. This is due to the high power required by one DAC per antenna. Those costs scale linearly with $N$ instead of with $K$, becoming prohibitive in the large array domain.

[5]Reasonably, $p(d_k)$ i.e. the probability to have LoS link, it is modeled with a monotonic non-increasing function of its argument.

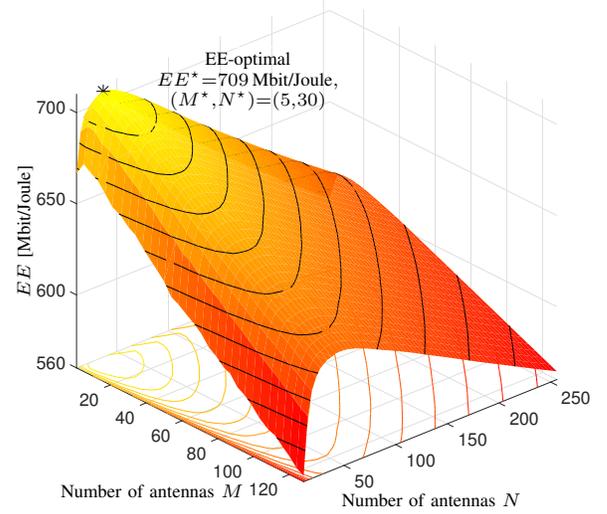

(a) Hybrid precoder

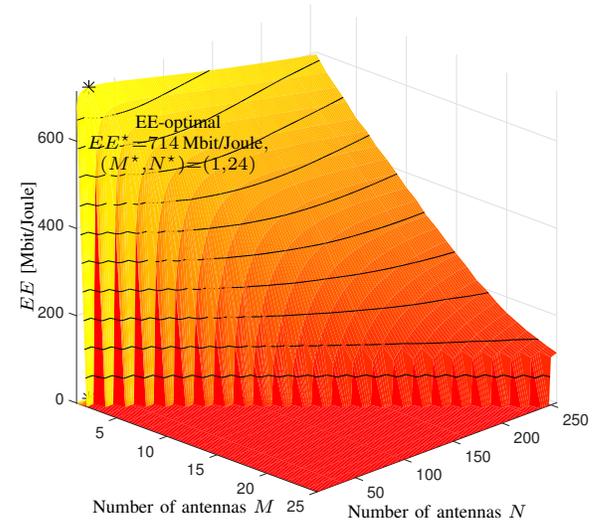

(b) Fully-digital precoder

Fig. 4: Energy Efficiency [Mbit/Joule] for different combination of $M$ and $N$ (with $K = 10$) for hybrid and fully-digital precoder.

TABLE II: Power consumption of the different components at the operating point $(M^\star, N^\star)$ with $P_{\text{FIX}} = 50W$.

| Power parameters | Hybrid | Fully-digital |
|---|---|---|
| $P_{\text{FIX}}$ | 68% | 17% |
| $P_{\text{RF}}$ | 5% | 16% |
| $P_{\text{FE}}$ | 24% | 65% |
| $P_{\text{LP}}$ | 3% | 2% |

## VII. CONCLUSIONS

This work focused on a two-tier network in which a given number $K$ of small-cell BSs uses a mmWave wireless backhaul to communicate with a macro BS. In particular, we analyzed how to select the number of BS antennas $N$ and number of receive antennas $M$ at each small-cell BS under the assumption that a hybrid transceiver architecture is employed, with the number of RF chains equal to $K$. To this end, we developed a realistic power consumption model that explicitly describes how the total power consumption of the hybrid scheme depends non-linearly on $M$, $K$, and $N$. Our analytical and numerical results showed that deploying a hybrid scheme with a large number of antennas $N$ is not the EE-optimal solution with today's technology. Alternative solutions must be developed in order to exploit the promising advantages (in terms of spectral efficiency) of using large values of $N$ at mmWave bands and at the same time to maximize the EE.